\documentclass[11pt]{article}
\usepackage[colorlinks=true]{hyperref}
\newcommand{\J}[1]{\mbox{${\cal I}_{#1}$}}
\newcommand{\fJ}[1]{\mbox{$\widetilde{\cal I}_{#1}$}}
\newcommand{\alp}{\alpha_{\scriptscriptstyle +}}
\newcommand{\alm}{\alpha_{\scriptscriptstyle -}}
\newcommand{\scsc}{\scriptscriptstyle}
\newcommand{\suma}{\sum_{\scsc k,\sigma}}
\newcommand{\loga}[1]{\log\!\left(#1\right)}
\newcommand{\logad}[1]{\log^2\!\left(#1\right)}
\newcommand{\zks}{z_{k\sigma}}
\newcommand{\yks}{y_{k\sigma}}
\newcommand{\zksk}{z_{k\sigma_k}}
\newcommand{\yksk}{y_{k\sigma_k}}
\newcommand{\yo}{y_{\scsc 0}}
\newcommand{\Sp}[1]{\mathrm{Li}_{2}\!\left(#1\right)}
\newcommand{\sgn}{\mathrm{sgn}}
\newcommand{\x}{\mathsf{x}}
\date{17 August 1999}
\title{Three and Two-Point One-Loop Integrals in Heavy Particle Effective
  Theories}
\author{Antonio O.\ Bouzas \thanks{E-mail:
    abouzas@casandra.cieamer.conacyt.mx}\\\small Departamento de F\'{\i}sica
  Aplicada, CINVESTAV-IPN \\\small Carretera Antigua a Progreso Km.\
  6, Apdo.\ Postal 73 ``Cordemex''\\\small
  M\'erida 97310, Yucat\'an, M\'exico}
\begin{document}
\maketitle
\begin{abstract}
  We give a complete analytical computation of three and
  two-point loop integrals occurring in heavy-particle
  theories, involving a velocity change, for arbitrary real values of
  the external masses and residual momenta.
\end{abstract}
\section{Introduction}
\label{sec:intro}
The study of the dynamics and spectroscopy of hadrons containing a
heavy quark has been greatly simplified and systematized with the
introduction of heavy quark effective theory (HQET) \cite{hqet}.
Heavy-particle theories along similar lines have also been succesfully
applied in other, related contexts.  Thus, in those cases where a
chiral approach to the strong interactions of heavy hadrons with light
mesons is applicable, a combination of chiral and heavy-quark
symmetries leads to heavy hadron chiral perturbation theory (HHChPT)
\cite{hhchpt}.  A heavy-particle expansion has also been developed in
the chiral-perturbative framework for nucleon-meson interactions,
which constitutes the so-called heavy baryon chiral perturbation
theory (HBChPT) \cite{hbchpt}.

In the heavy-quark limit the interaction of a heavy quark, or hadron,
with the light degrees of freedom cannot change its four-velocity
$v^\mu$.  In consequence, $v^\mu$ becomes a good quantum number and,
therefore, heavy-particle effective theories of the strong
interactions are expressed in terms of velocity-dependent fields.
Weak interactions, or other external sources, however, can change the
velocity and/or flavor of a heavy quark or hadron.  Strong-interaction
corrections to velocity-changing interaction vertices then involve
loop integrals with two different velocities.

In this paper we report on a complete analytic computation of
three-point loop integrals involving a velocity change, and two-point
loop integrals.  We consider a class of one-loop integrals occurring
in heavy-particle theories, with arbitrary real values for the
external masses and residual momenta.  Since our aim here is mainly
methodological, we will not discuss specific phenomenological
applications.  For definiteness, however, we adopt the language of
HHChPT in the sequel.

In the next section, we define the integrals to be studied, establish
our notations and conventions, and discuss the method we use, which
involves a combination of the HQET technique and of standard methods
for computing loop integrals \cite{thooft,passar,denner}. In section 3
we give technical details about the computation of the scalar
three-point integral, state our results and discuss several important
limits and particular cases and cross-checks.  In section 4 we briefly
consider the two-point integral, which has already been given in the
previous literature.  In section 5 the vector and second-rank tensor
integrals are given in terms of form factors.  Finally, in section 6
we give some final remarks.

\section{Method. Notation and Conventions}
\label{sec:method}

The loop integrals we consider are of the form, 
\begin{eqnarray} 
  \J{3}^{\alpha_1\cdots \alpha_n} & = & \frac{i\mu^{4-d}}{(2\pi)^d}
  \int\! d^d\! q \frac{q^{\alpha_1}\cdots q^{\alpha_n}}{
    \left(2v_1\!\cdot\! (q+k_1)-\delta M_1+i\varepsilon\right)
    \left(2v_2\!\cdot\! (q+k_2)-\delta M_2+i\varepsilon\right)
    \left(q^2-m^2+i\varepsilon\right)}  \label{j3}\\
  \J{2}^{\alpha_1\cdots \alpha_n} & = & \frac{i\mu^{4-d}}{(2\pi)^d}
  \int\! d^d\! q \frac{q^{\alpha_1}\cdots q^{\alpha_n}}{
    \left(2v\!\cdot\! (q+k)-\delta M+i\varepsilon\right)
    \left(q^2-m^2+i\varepsilon\right)}.  \label{j2}
\end{eqnarray}
Here $v_i^\mu$, $i=1,2$, are the velocities of the external heavy
legs, $k_i^\mu$ their residual momenta, and $\delta M_i$ their mass
splittings relative to the common heavy mass of the corresponding
heavy-flavor/spin multiplet.  $m$ is the mass of the light particle,
which in HHChPT corresponds to a light pseudoscalar meson.  These
integrals are defined in $d=4-\epsilon$ dimensions, $\mu$ being the
mass scale of dimensional regularization.  Their degrees of divergence
are $n+d-4$ for $\J{3}^{\alpha_1\cdots \alpha_n}$ and $n+d-3$ for
$\J{2}^{\alpha_1\cdots \alpha_n}.$ The factor of 2 in front of
$v_i^\mu$ corresponds to our normalization of the heavy-particle
propagators.

In this section we will restrict ourselves to the scalar case $n=0$.
The cases $n=1,2$ will be considered in detail in section 5.  Together
with $\J{2,3}$ we consider also the auxiliary integrals,
\begin{eqnarray} 
  \fJ{3} & = & \frac{i\mu^{4-d}}{(2\pi)^d}
  \int\! d^d\! q \frac{1}{
    \left((q+p_1)^2- M_1^2+i\varepsilon\right)
    \left((q+p_2)^2- M_2^2+i\varepsilon\right)
    \left(q^2-m^2+i\varepsilon\right)}  \label{fj3}\\
  \fJ{2} & = & \frac{i\mu^{4-d}}{(2\pi)^d}
  \int\! d^d\! q \frac{1}{
    \left((q+p)^2-M^2+i\varepsilon\right)
    \left(q^2-m^2+i\varepsilon\right)}.  \label{fj2}
\end{eqnarray}
$\fJ{3}$ is convergent in four dimensions, with degree of divergence
$d-6$. $\fJ{2}$ has degree of divergence $d-4$, diverging
logarithmically at $d=4$.  The relations among external momenta
and masses in $\J{3}$ and $\fJ{3}$ are, $(i=1,2)$
\begin{equation} 
  \label{param1}
  p_i^\mu  =  M v_i^\mu + k_i^\mu~,
  \hspace{2ex}
  p_i^\mu p_{i\mu} > 0~;
  \hspace{3ex}
  M_i  =  M+\frac{1}{2} \delta M_i~,
  \hspace{2ex}
  M_i > 0, 
\end{equation}
and similarly for $\J{2}$ and $\fJ{2}.$ We remark at this point that
$\fJ{2,3}$ need not be related to Feynman diagrams in any existing
physical theory.  The similarity of the limit $M\rightarrow\infty$ 
studied below with the heavy-quark limit is purely formal.
$\fJ{2,3}$ are just intermediate steps in our calculation of
$\J{2,3}$, as we discuss next.

In the limit $M\rightarrow\infty$ we have, ($k=1,2$)
\begin{equation}
  \label{hprop}
  \frac{1}{(q+p_k)^2- M_k^2+i\varepsilon} =
  \frac{1}{M} \frac{1}{2v_k\!\cdot\! (q+k_k)-\delta
  M_k+i\varepsilon}
  + {\cal O}\left(\frac{1}{M^2}\right).
\end{equation}
Since $\partial \J{2,3}/\partial m^2$ are convergent for
$d=4$, equation (\ref{hprop}) leads to,
\begin{equation}
  \label{matchm}
  \frac{\partial \J{3}}{\partial m^2} = M^2 \frac{\partial
  \fJ{3}}{\partial m^2} + {\cal O}\left(\frac{1}{M}\right),
\hspace{3em}
  \frac{\partial \J{2}}{\partial m^2} = M\frac{\partial
  \fJ{2}}{\partial m^2} + {\cal O}\left(\frac{1}{M}\right)~.
\end{equation}
Therefore, at $d=4$ we must have, 
\begin{eqnarray}
  \label{matchm2}
  \J{3} &=& \J{3}|_{\scriptscriptstyle m=0} + M^2
  \left(\fJ{3}-\fJ{3}|_{\scriptscriptstyle m=0}\right) + 
  {\cal O}\left(\frac{1}{M}\right) \\
   \J{2} &=& \J{2}|_{\scriptscriptstyle m=0} + M
  \left(\fJ{2}-\fJ{2}|_{\scriptscriptstyle m=0}\right) + 
  {\cal O}\left(\frac{1}{M}\right).   \label{matchm3}
\end{eqnarray}
Moreover, using equations (\ref{param1}) and defining $\Delta_j \equiv
\delta M_j -2 v_j\! \cdot\! k_j$, we can write $\J{3}$, $\fJ{3}$ in
terms of $\Delta_i.$ Differentiating we obtain, to leading order in
$1/M$,
\begin{equation}
  \label{diffd}
  \frac{\partial \J{3}}{\partial \Delta_j} = M^2 \frac{\partial
  \fJ{3}}{\partial \Delta_j} + {\cal O}\left(\frac{1}{M}\right), 
\hspace{5em}  j=1,2,
\end{equation}
or, equivalently,
\begin{equation}
  \label{equiv}
  \J{3} = M^2 \fJ{3}+C_1(\Delta_1)+ {\cal O}\left(\frac{1}{M}\right)=
          M^2 \fJ{3}+C_2(\Delta_2)+ {\cal O}\left(\frac{1}{M}\right)~.
\end{equation}
Here, the dependence of $C_{1,2}$ on $d, M, \mu, m$ and $ v_1\!\cdot\!v_2$
is understood, but it is shown explicitly that $C_1$ can depend on
$\Delta_1$ but not on $\Delta_2$, and the oposite is true for $C_2.$
Subtracting the two equations (\ref{equiv}) term by term, we conclude
that $C_1=C_2=C(d, M, \mu, m, v_1\!\cdot\!v_2)$ do not depend on
$\Delta_{1,2}$.  (Furthermore, (\ref{matchm}) together with
(\ref{diffd}) imply that $C$ does not depend on $m$ either.) Thus, 
\begin{equation}
  \label{matchd}
    \J{3} = \J{3}|_{\scriptscriptstyle \Delta_1=0=\Delta_2} + M^2
  \left(\fJ{3}-\fJ{3}|_{\scriptscriptstyle \Delta_1=0=\Delta_2}\right)
  +{\cal O}\left(\frac{1}{M}\right)~.
\end{equation}
We notice that $\J{3}$ is straightforward to compute for
$\Delta_j=0$ by using the HQET method for combining
denominators (see, \emph{e.g.,} \cite{cho}). On the other hand, $\fJ{3}$
is needed in (\ref{matchd}) only at $d=4$, and to leading order in
$M$, including logarithmic corrections.  Eq.\ (\ref{matchd}) will
then be the starting point for our computation of $\J{3}$. Equations
for $\J{2}$ analogous to (\ref{equiv}) and (\ref{matchd}) can also be
obtained, involving two derivatives. We will find it more convenient
to use eq.\  (\ref{matchm3}) in order to compute $\J{2}.$

Scalar integrals can depend on $v_1^\mu$, $v_2^\mu$ only through
$\omega=v_1\cdot v_{2}$.  If we denote by $\Omega$ the
magnitude of the three-velocity associated to $v_1^\mu$ or $v_2^\mu$
in the rest frame of $v_1^\mu + v_2^\mu$ then \cite[\S 11.5]{jackson},
\begin{equation}
  \label{omega}
  \omega=v_1\!\cdot\! v_{2}=\frac{1+\Omega^2}{1-\Omega^2}~;
  \hspace{5ex}
  \Omega=\sqrt{-\frac{(v_1^\mu-v_2^\mu)^2}{(v_1^\mu+v_2^\mu)^2}} =
  \sqrt{\frac{\omega-1}{\omega+1}}~. 
\end{equation}
Together with $\Omega$, the roots of $(v_1^\mu-\alpha v_2^\mu)^2=0$, given
by
\begin{equation}
  \label{alpm}
  \alpha_{\scsc\pm} =
  \omega\pm\sqrt{\omega^2-1}=\frac{1\pm\Omega}{1\mp\Omega}~, 
\end{equation}
will appear frequently below.  For physical values of
$v_{1,2}^\mu$, such that $(v_{1,2}^{\mu})^2=1$, we have $\omega>1$,
$0<\Omega<1$, $0<\alm<1<\alp$.  We will always assume these
inequalities  to hold in what follows. 

Logarithms have a cut along the negative real axis.  The log of a
product can be split as $\loga{ab}=\loga{a}+\loga{b}$ if
$\mathrm{Im}(a)$ and $\mathrm{Im}(b)$ have opposite sign, or if $a>0$.
Similarly, $\loga{a b}=\loga{a}-\loga{b}$ if $\mathrm{Im}(a)$ and
$\mathrm{Im}(b)$ have the same sign, or if $a>0$ \cite{thooft}.  Given
a complex number $z$, we use that determination of the argument such
that $-\pi < \arg(z) < \pi$.  In particular, $\loga{1/z} = -\loga{z}$.
We use the same definition and conventions as \cite{thooft} for the
dilogarithm, which we denote by $\mathrm{Li}_2$.

\section{The Scalar Three-Point Integral}
\label{sec:3scalar}

We will now consider in detail the calculation of $\J{3}.$ Our first
step is to compute $\fJ{3}$ to leading order in $M.$ As mentioned
above, we only need to evaluate $\fJ{3}$ at $d=4$.  We introduce a
standard Feynman parametrization of the integrand in (\ref{fj3}).
Integrating over $d^4q$ and over the Feynman parameter associated with
the third propagator in (\ref{fj3}), we obtain,
\begin{equation}
  \label{fj3a}
  \fJ{3} = \frac{1}{(4\pi)^2} \int_0^1\! dx\int_0^{1-x}\! dy 
\frac{1}{(yp_1+xp_2)^2+y (M_1^2-p_1^2-m^2)+x
  (M_2^2-p_2^2-m^2)+m^2-i\varepsilon}. 
\end{equation}
In the limit $M\rightarrow\infty$ the polynomial in the denominator is
a sum of terms of the form $M^p x^q y^{p-q}$, with $p=0,1,2$ and
$0\leq q\leq p$, the term with $p=0$ being $m^2-i \varepsilon$.  These
are properties we want to maintain, in order to be able to take the
limit $M\rightarrow\infty$ later, retaining only the leading terms in
$M$ in each coefficient.  Thus, we will not make a change of variable
$y\rightarrow 1-y$ at this stage, as the limits of integration
suggest.  Following \cite{thooft}, we shift variables according to
$y\rightarrow y-\alpha x$.  This shift is homogeneous in $x$, $y$, so
it does not change the order of each term as $M\rightarrow\infty$.
The parameter $\alpha$ is taken to be one of the roots $\alpha_{\pm}$
of $(p_2 -\alpha p_1)^2=0.$ For $M$ large, we have $\alp>1>\alm>0$.
Choosing $\alpha=\alp$ and exchanging the order of integration, we are
led to,
\begin{eqnarray}
  \label{fj3b}
  \fJ{3}  & = & \frac{1}{(4\pi)^2} \left\{\int_0^1\!
    dy\int_0^{y/\alp}\! dx + \int_1^{\alp}\!
    dy\int_{\frac{y-1}{\alp-1}}^{y/\alp}\! dx 
  \right\} \frac{1}{D}\\
 D & = & \left(2y(p_1\!\cdot\!p_2-\alp p_1^2) + (M_2^2-p_2^2-m^2) -
   \alp (M_1^2-p_1^2-m^2)\right) x \nonumber\\
 & & +p_1^2 y^2+y
    (M_1^2-p_1^2-m^2)+m^2-i\varepsilon~. \nonumber
\end{eqnarray}
Using equations (\ref{param1}) and retaining only
leading powers of $M$ in each coefficient, the previous expression
simplifies considerably.  In the limit $M\rightarrow\infty$,
$\alpha_{\pm}$ are given by (\ref{alpm}).  Performing the integration
over $x$ we obtain,
\begin{eqnarray} 
  \label{fj3c}
  \fJ{3} & = & -\frac{1}{64\pi^2} \frac{1-\Omega^2}{\Omega} \frac{1}{M^2}
  \int_0^1\! dy \frac{1}{G} \loga{\frac{K}{H}}-
  \frac{1}{64\pi^2} \frac{1-\Omega^2}{\Omega} \frac{1}{M^2}
  \int_1^{\alp}\! dy \frac{1}{G} \loga{\frac{K}{L}}\\
  L & = & H -\frac{4\Omega}{1-\Omega^2}\, \frac{y-1}{\alp-1}\, G~;
  \hspace{3ex}
  K  = H -\frac{4\Omega}{1-\Omega^2}\, \frac{y}{\alp}\, G \nonumber\\
  H & = & y^2+\frac{\Delta_1}{M}\, y +\frac{m^2}{M^2}-i\varepsilon ~;
  \hspace{3ex}
  G = y-\frac{\yo}{M} \nonumber~,
\end{eqnarray}
with
\begin{equation}
  \label{yodd}
  \yo  =  -\frac{1+\Omega}{2\Omega} (\Omega\Delta+\delta)~;
  \hspace{3ex}
  \Delta  =  \frac{1}{2} (\Delta_1+\Delta_2) ~;
  \hspace{3ex}
  \delta  =  \frac{1}{2} (\Delta_1-\Delta_2) ~.
\end{equation}
In these last two equations we have introduced several notations
that will be needed later.  In the second integral in (\ref{fj3c}) the
variable $y$ is ${\cal O}(1)$ over the entire domain of integration.
Therefore, $My={\cal O}(M)$ and the integral is given, to leading
order in $M$, by
\begin{equation}
  \label{fj3d}
  -\frac{1}{64\pi^2} \frac{1-\Omega^2}{\Omega} \frac{1}{M^2}
  \int_1^{\alp}\! dy\, \frac{1}{y}\,
  \log\!\left(\alm \frac{y-i\varepsilon}{-y+\alp+1-i\varepsilon}\right)~. 
\end{equation}
This expression does not depend on $\Delta_{1,2}$.  Therefore, it will
cancel when we subtract $\fJ{3}|_{\scriptscriptstyle
  \Delta_1=0=\Delta_2}$ from $\fJ{3}$, and will not contribute to
$\J{3}$ as given in (\ref{matchd}).  We shall then drop this term from
$\fJ{3}$ from now on.  The remaining integral can be re-written as,
\begin{equation}
  \label{fj3e}
  \fJ{3}  =  \frac{1}{64\pi^2} \frac{1-\Omega^2}{\Omega}
  \frac{1}{M^2} \int_0^1\! dy\, \frac{1}{G}
  \left\{\loga{H}-\loga{K}\right\} + \cdots~.
\end{equation}
We notice that there is no singularity at the zero of the denominator,
since the numerator vanishes there.  We will denote $y_{1\pm}/M$ and
$y_{2\pm}/M$ the roots of $H$ and $K$, respectively.  They are given
by, 
\begin{equation}
  \label{roots}
  y_{\scsc 1\pm} = \frac{1}{2} \left(-\Delta_1\pm\sqrt{\Delta_1^2-4
    m^2+i\varepsilon}\right)~; 
  \hspace{5ex}
  y_{\scsc 2\pm} = \frac{\alp}{2} \left(-\Delta_2\pm\sqrt{\Delta_2^2-4
    m^2+i\varepsilon}\right)~. 
\end{equation}
From their definition, (\ref{fj3c}), it is clear that $H$ and $K$ are
equal at the zero of $G$.  Defining, 
\begin{equation}
  \label{zks}
  \zks = \yks -\yo~,
  \hspace{5ex}
  k=1,2, ~~~\sigma=\pm~, 
\end{equation}
the equality of $H$ and $K$ at $y=\yo/M$ can be expressed as,
\begin{equation}
  \label{iden}
  z_{\scsc 1+} z_{\scsc 1- } = \alm^2 z_{\scsc 2+} z_{\scsc 2-}~,
\end{equation}
an identity that will be important below.

After factorizing $H$ and $K$ and splitting the logs in (\ref{fj3e}),
we find, 
\begin{eqnarray}
  \label{fj3f}
  \fJ{3} & = & \frac{1}{64\pi^2 }\frac{1-\Omega^2}{\Omega}
  \frac{1}{M^2}
  \int_0^1\! dy\,
  \frac{1}{y-\yo/M} \left\{
    \log\!\left(y-\frac{y_{\scsc
  1+}}{M}\right)+\log\!\left(y-\frac{y_{\scsc 1-}}{M}\right)\right.
  \nonumber\\
  & & \left. - \log\!\left[\alm\left(y-\frac{y_{\scsc 2+}}{M}\right)\right]
  - \log\!\left[\alm\left(y-\frac{y_{\scsc 2-}}{M}\right)\right]\right\}~.
\end{eqnarray}
In order to be able to distribute the integral inside the braces
without introducing spurious singularities, we use (\ref{iden}) to add
and subtract the value of each log at the pole.  In this way we~obtain,
\begin{equation}
  \label{fj3g}
  \fJ{3}  =  \frac{1}{64\pi^2} \frac{1-\Omega^2}{\Omega} \frac{1}{M^2}
  \suma (-1)^{k+1} \int_0^1\!
 dy\,\frac{1}{y-\yo/M}
 \left\{\log\!\left(y-\frac{y_{k\sigma}}{M}\right)
   -\log\!\left(\frac{\yo}{M}-\frac{y_{k\sigma}}{M}\right)\right\} ~, 
\end{equation}
where the sum runs over $k=1,2$ and $\sigma=\pm$.
These integrals are already in standard form.  Evaluating them to
leading order in $M$, we arrive~at,
\begin{eqnarray}
  \label{fj3h}
  \fJ{3} & = & \frac{1}{64\pi^2} \frac{1-\Omega^2}{\Omega}
  \frac{1}{M^2}
  \suma (-1)^{k+1} \left\{\frac{1}{2} \logad{\frac{M}{\mu}}
    -\frac{\pi^2}{6}
    -\loga{\frac{M}{\mu}} \loga{-\frac{\zks}{\mu}}\right.\nonumber\\
    & & \left. -\frac{1}{2} \logad{\frac{\zks}{\mu}}
    + \loga{\frac{\zks}{\mu}} \loga{-\frac{\zks}{\mu}}
    -\loga{-\frac{\yo}{\zks}} \loga{\frac{\yks}{\zks}}
    -\Sp{\frac{\yks}{\zks}}\right\}~.
\end{eqnarray}
In order to simplify this result we have explicitly used the relation
$\log(\mu/\zks)=-\log(\zks/\mu)$ as explained in section
\ref{sec:method}.  Notice that we have introduced a mass scale $\mu$
that makes the arguments of the logs dimensionless.  The first two
terms in (\ref{fj3h}) are independent of $k$ and therefore they cancel
out in the sum.  The third term can be simplified by making use of
(\ref{iden}), which results in a term of the form 
$(1-\Omega^2)/(64\pi^2 M^2\Omega) \log(\alp^2) \log(M/\mu)$.

The integral $\fJ{3}$ at $\Delta_j=0$ is straightforward
to compute.  Its logarithmic dependence on $M$ cancels that of
$\fJ{3}$, so that we obtain,
\begin{eqnarray}
  \label{fj3i}
  \fJ{3}-\fJ{3}|_{\scsc \Delta_j=0} & = &
  \frac{1}{32\pi^2 M^2} \frac{1-\Omega^2}{2\Omega}
  \left\{ \logad{\alp} + \loga{\alp^2} \loga{\frac{m}{\mu}}
  + \suma
  (-1)^{k} \left[\frac{1}{2} \logad{\frac{\zks}{\mu}}\right.\right.\nonumber\\
& & \left.\left. - \loga{\frac{\zks}{\mu}} \loga{-\frac{\zks}{\mu}}
    +\loga{-\frac{\yo}{\zks}} \loga{\frac{\yks}{\zks}}
    +\Sp{\frac{\yks}{\zks}}\right]\right\}~.
\end{eqnarray}
This is essentially the final result, except for the dimensional
regularization pole term, which is supplied by $\J{3}|_{\scsc
  \Delta_j=0}$ (see eq.\ (\ref{matchd})). 
Setting $n=0$, $\delta M_j=0=k_j$, $j=1,2$, in (\ref{j3}), we obtain,
after using the HQET method for combining denominators and integrating
over $d^dq$,
\begin{equation}
  \label{j3a}
  \J{3}|_{\scsc \Delta_j=0} = \frac{\mu^{4-d}}{(4\pi)^{d/2}}\,
  \Gamma\!\left(3-\frac{d}{2}\right) \int_0^1\!dx \int^\infty_0\! d\lambda 
  \frac{\lambda}{\left[\lambda^2 \left(x^2+(1-x)^2+2 x(1-x) \omega\right)+
    m^2-i\varepsilon\right]^{3-d/2}}~.
\end{equation}
The innermost integral can be evaluated by changing variable to
$u=\lambda^2$. We get,
\begin{eqnarray}
\J{3}|_{\scsc \Delta_j=0} & = & \frac{1}{2 (4\pi)^{d/2}}
\Gamma\left(2-\frac{d}{2}\right) \left(\frac{\mu}{m}\right)^{4-d}
\int_0^1\! dx \frac{1}{x^2+(1-x)^2+2 x (1-x) \omega}  \label{j3b}\nonumber\\
& = & \frac{1}{2 (4\pi)^{d/2}}
\Gamma\left(2-\frac{d}{2}\right) \left(\frac{\mu}{m}\right)^{4-d}
\frac{1-\Omega^2}{4\Omega} \log\!\left(\alp^2\right)   \label{j3c}\nonumber\\
& = & \frac{1}{64\pi^2} \frac{1-\Omega^2}{\Omega} \loga{\alp} \left(
  \frac{2}{\epsilon}-\gamma_{\scsc\mathrm{E}}
  +\loga{4\pi}+\loga{\frac{\mu^2}{m^2}}\right) + {\cal O}
(\epsilon)  \label{innermost}~,
\end{eqnarray}
where $\gamma_{\scsc\mathrm{E}}$ is Euler's gamma.  

Thus, finally, using (\ref{matchd}), (\ref{fj3i}) and (\ref{innermost}), we
obtain,
\begin{eqnarray}
  \lefteqn{
  \J{3}  =  \frac{1}{64\pi^2} \frac{1-\Omega^2}{\Omega}
  \left\{\frac{2}{\epsilon} \loga{\alp} +\logad{\alp}
    +\suma (-1)^k \left[\frac{1}{2} \logad{\frac{\zks}{\overline\mu}}
  \right.\right. }\nonumber\\
  & & \left. \left. 
      -\loga{\frac{\zks}{\overline\mu}}
      \loga{-\frac{\zks}{\overline\mu}} +\loga{-\frac{\yo}{\zks}}
      \loga{\frac{\yks}{\zks}} +\Sp{\frac{\yks}{\zks}}
    \right]\rule{0cm}{4ex}\right\} ~,\label{finally}
\end{eqnarray}
where we used (\ref{iden}) to rewrite $\log(\mu)$ in (\ref{fj3i}) as
$\log(\overline\mu) = \log(\mu\sqrt{4\pi}) -
\gamma_{\scsc\mathrm{E}}$.  Equation (\ref{finally}) is our main
result.  It gives the analytical expression for $\J{3}$ for general
real values of the external masses and residual momenta.  Below we
consider some particular values of the parameters which are
important in practice, and in which the expression for $\J{3}$ takes
on a simplified form.  They can also serve as cross-checks of
(\ref{finally}).

We note that $\J{3}$, given by (\ref{j3}) with $n=0$, is symmetric
under exchange of heavy particles $1\leftrightarrow 2$. The general
result (\ref{finally}) is not manifestly invariant under
$\Delta_1\leftrightarrow \Delta_2$, but its symmetry has been
thoroughly checked numerically.

\subsection{The case $\yo<0$ and the zero-recoil limit}
\label{sec:zrecoil}

An important particular case to consider is the case $v_1^\mu=v_2^\mu$
or, equivalently, $\Omega=0$ ($\omega=1$).  $\J{3}$ at $\Omega=0$ can
be computed more directly by differentiating $\J{2}$ (see section
\ref{sec:2scalar}). Therefore, its calculation from (\ref{finally})
constitutes a cross-check.

The limit $\Omega\rightarrow 0$, however, is difficult to take in
(\ref{finally}).  As indicated by the factor $1/\Omega$ in
that equation, the limit results from complicated cancellations
among different terms in the sum in (\ref{finally}).  Individually,
some of those terms may be large as $\Omega\rightarrow 0$.  This
situation arises mainly from the addition and subtraction of terms of
the form $\log(\yo/M-\yks/M)$ in (\ref{fj3g}), which is necessary when
$\yo>0$, since in that case there is a singularity in the integration
domain.  If, however, we restrict ourselves to the case $\yo<0$, then
the general expression (\ref{finally}) takes a simpler form, which
makes the zero-recoil limit transparent. 

As mentioned at the end of the previous section, $\J{3}$ is symmetric
under exchange of $\Delta_1$ and $\Delta_2$.  Hence, without loss of
generality we can assume $\delta>0$ (see eq.\ (\ref{yodd}). The case
$\delta=0$ will be considered~afterwards as a limiting case).  From
the definition of $\yo$ in (\ref{yodd}) we see that, if $\delta>0$,
then for sufficiently small $\Omega$ we will have $\yo<0$.  This is
therefore the relevant parameter region to consider when
$\Omega\rightarrow 0$.

Assuming, then, $\yo<0$, we can go back to (\ref{fj3f}) and distribute
the integral inside the braces without adding extra terms.  The
calculation goes through unchanged, yielding,
\begin{eqnarray}
\label{j3y0<0}
\left.\J{3}\rule{0ex}{2ex}\right|_{\stackrel{\scsc \delta>0}{\scsc \yo<0}} & = &
\frac{1}{64\pi^2} \frac{1-\Omega^2}{\Omega} \left\{\rule{0ex}{4ex}
  \loga{\alp} \left[ \frac{2}{\epsilon} + \loga{\alp} +
  \loga{\frac{\overline\mu^2}{4\Omega^2 \yo^2}} \right]\right.\nonumber\\
& & + \left.\suma (-1)^k
\left[ \frac{1}{2} \logad{1-\frac{\yks}{\yo}} +
  \loga{1-\frac{\yks}{\yo}}
  \loga{\frac{2\Omega\yo}{\yks}}+\Sp{\frac{\yks}{\zks}}\right] 
\rule{0ex}{4ex}\right\}~.
\end{eqnarray}
The zero-recoil limit of this expression can be easily obtained.
Using the assumption $\delta>0$ and the relation (valid for $\Omega=0$)
\[\suma (-1)^k \yks = 2\delta~,\]
we find,
\begin{equation}
  \label{j3zr}
  \left.\J{3}\rule{0ex}{2ex}\right|_{\stackrel{\scsc \delta>0}{\scsc
      \Omega=0}}  = 
  \frac{1}{32\pi^2} \left(\frac{2}{\epsilon} + 2 + \suma (-1)^{k+1}
      \frac{\yks}{\delta} \loga{-\frac{\yks}{\overline\mu}}\right)~.
\end{equation}
It is understood that in this equation we must set $\Omega=0$ in the
expression (\ref{roots}) for $\yks$.  In section \ref{sec:2scalar} we
will obtain this result from $\J{2}$.

We give, finally, the result for $\J{3}$ at zero-recoil when
$\delta=0$ ($\Delta_1 = \Delta_2 \equiv \Delta$).  We just take the limit
of (\ref{j3zr}) as $\delta\rightarrow 0$ to obtain,
\begin{equation}
  \label{j3zrd0}
  \left.\J{3}\rule{0ex}{2ex}\right|_{\stackrel{\scsc \delta=0}{\scsc
      \Omega=0}}  =    \frac{1}{32\pi^2} \left\{\frac{2}{\epsilon}
    +\loga{\frac{\overline\mu^2}{m^2}} + \frac{\Delta}{\sqrt{\cdot}}
    \left[\loga{\frac{\Delta-\sqrt{\cdot}}{\overline\mu}} -
    \loga{\frac{\Delta+\sqrt{\cdot}}{\overline\mu}}  \right]
  \right\}~,
\end{equation}
where $\sqrt{\cdot}=\sqrt{\Delta^2-4 m^2+i\varepsilon}$.  Notice that,
in general, we cannot express the difference of logs in (\ref{j3zrd0})
as the log of the ratio of the arguments , because their imaginary
parts have opposite sign.

\subsection{The case $\yo=0$}
\label{sec:yo0}

When $\yo=0$ the general expression (\ref{finally}) for $\J{3}$ is
singular.  The singularity is avoidable, though, so that we can take
the limit $\yo\rightarrow 0$ in (\ref{finally}) to get,
\begin{eqnarray}
  \label{yo0}
  \left.\J{3}\rule{0ex}{2ex}\right|_{\scsc \yo=0} & = &
  \frac{1}{64\pi^2} \frac{1-\Omega^2}{\Omega} 
    \loga{\alp} \left(\frac{2}{\epsilon}+\loga{\alp}\right)\nonumber\\
  & &   + \frac{1}{64\pi^2} \frac{1-\Omega^2}{\Omega} \suma (-1)^k
  \left[\frac{1}{2}\logad{\frac{\yks}{\overline\mu}} 
    -\loga{\frac{\yks}{\overline\mu}}
  \loga{-\frac{\yks}{\overline\mu}}\right] ~.
\end{eqnarray}
This expression is valid, in particular, when $\Delta_1 =0=\Delta_2$,
which is the point in parameter space we used to ``match'' $\J{3}$ and
$\fJ{3}$.  As is easily seen, in that case we recover
(\ref{innermost}).

\subsection{The case $m=0$}
\label{sec:m0}

The case $m=0$ is relevant to theories involving massless particles,
such as gluons in HQET and Goldstone bosons in chiral theories in the
limit of massless quarks .  The value of $\J{3}$ at $m=0$ can be
obtained from the general expression (\ref{finally}), and also by
direct computation from (\ref{j3}) using the HQET method for combining
denominators.  We will consider both approaches in this section.
Together with the zero-recoil case studied in sections
\ref{sec:zrecoil} and \ref{sec:2scalar}, this is one of our main
cross-checks.

We will now study the limit $m\rightarrow 0$ of $\J{3}$ as given in
(\ref{finally}).  We notice that when $m=0$ one of the two possible
values $\sigma=\pm$ leads to $\yks=0$ (see eq.\ (\ref{roots})).  In
fact, $m=0$ and $\Delta_k > 0$ (resp.\ $\Delta_k < 0$) implies
$y_{k+}=0$ (resp.\ $y_{k-}=0$). Calling
$\sigma_k=-\sgn(\Delta_k)=-\sigma_k^\prime$, so that
$y_{k\sigma_k}\neq 0 = y_{k\sigma_k^\prime}$, we have
$z_{k\sigma_k^\prime}= -\yo+i\varepsilon \sigma_k^\prime$ and,
therefore,
\[\loga{-\frac{\yo}{z_{k\sigma_k^\prime}}}
\loga {\frac{y_{k\sigma_k^\prime}}{z_{k\sigma_k^\prime}}}=0~;
\hspace{5ex}
\Sp{\frac{y_{k\sigma_k^\prime}}{z_{k\sigma_k^\prime}}}=0~,\] 
so these terms drop from the sum in (\ref{finally}).  On the other
hand, since $\yo$ is by definition (\ref{yodd}) a real number, we have
the equalities,
\begin{eqnarray*}
  \sum_{k=1}^2 (-1)^k \frac{1}{2}
  \logad{-\yo+i\varepsilon\sigma^\prime_k} & = & i\pi \theta(\yo)
  \loga{\yo} \sum_{k=1}^2 (-1)^k \sigma^\prime_k\\
  \sum_{k=1}^2 (-1)^k 
  \loga{-\yo+i\varepsilon\sigma^\prime_k}
  \loga{\yo-i\varepsilon\sigma^\prime_k}& = & i\pi \sgn(\yo)
  \loga{|\yo|} \sum_{k=1}^2 (-1)^k \sigma^\prime_k~,
\end{eqnarray*}
$\theta(x)$ being a step function.  With these relations taken into
account, from (\ref{finally}) we find for $\J{3}$,
\begin{eqnarray}
  \left.\J{3}\rule{0ex}{2ex}\right|_{\scsc m=0} & = &
  \frac{1}{64\pi^2} \frac{1-\Omega^2}{\Omega}
  \left\{\frac{2}{\epsilon} \loga{\alp} +\logad{\alp}+\sum_{k=1}^2
    (-1)^k \left[\rule{0ex}{4ex}\frac{1}{2}
      \logad{\frac{\zksk}{\overline\mu}}\right.\right.\nonumber\\
  & & \left. \left.  -\loga{\frac{\zksk}{\overline\mu}}
      \loga{-\frac{\zksk}{\overline\mu}} +\loga{-\frac{\yo}{\zksk}}
      \loga{\frac{\yksk}{\zksk}} +\Sp{\frac{\yksk}{\zksk}} + {\cal
        E}(\yo) \rule{0ex}{4ex}\right]\right\},  \label{m=0}\\
  \label{cale}
  {\cal E}(\yo)  & \equiv &  i\pi \theta(-\yo)
  \loga{\frac{|\yo|}{\overline\mu}} \sum_{k=1}^2 (-1)^k
  \sigma_k^\prime
   =  -2i\pi \loga{\frac{-\yo}{\overline\mu}}
  \theta(\Delta_1)\theta(-\Delta_2)~.  
\end{eqnarray}
Notice that the sum in (\ref{m=0}) runs over $k=1,2$ but not over
$\sigma_k$, and that $-\yo>0$ in (\ref{cale}) due to the step
functions.  As before, $\yksk$ and $\zksk$ in (\ref{m=0}) are given by
(\ref{roots}) and (\ref{zks}) with $m$ set to zero.
When $\delta=0$ ($\Delta_1=\Delta_2\equiv\Delta$), equation
(\ref{m=0}) becomes
\begin{eqnarray}
  \left.\J{3}\rule{0ex}{2ex}\right|_{\scsc m=0=\delta} & = &
  \frac{1}{64\pi^2} \frac{1-\Omega^2}{\Omega}
  \left\{\frac{2}{\epsilon} \loga{\alp} -\frac{1}{2} \loga{\alp}
    \loga{\frac{1-\Omega^2}{4}}-2 \loga{\alp}
  \loga{\frac{\Delta-i\varepsilon}{\overline\mu}}
 \right.\nonumber\\
 &  & \left. + \Sp{\frac{1-\Omega}{2}} - \Sp{\frac{1+\Omega}{2}}
 \rule{0ex}{4ex}\right\}~,  \label{m=0=d}
\end{eqnarray}
a result we will explicitly cross-check below.

We now turn to the calculation of $\J{3}$ at $m^2=0$ directly from its
definition (\ref{j3}) with the HQET method for combining
denominators.  For the sake of brevity, we will skip the details of
the derivation and quote the final result, which can be written as,
\begin{eqnarray}
  \lefteqn{
  \left.\J{3}\rule{0ex}{2ex}\right|_{\scsc m=0}  = 
  \frac{1}{64\pi^2} \frac{1-\Omega^2}{\Omega}
  \left\{\frac{2}{\epsilon} \loga{\alp} -\frac{1}{2} \loga{\alp}
  \loga{\frac{1-\Omega^2}{4}} + \Sp{\frac{1-\Omega}{2}} -
  \Sp{\frac{1+\Omega}{2}} \right\}
  } \nonumber\\
& & + \frac{1}{64\pi^2} \frac{1-\Omega^2}{\Omega} \left\{
  -\loga{\frac{-(1+\Omega)\delta}{\Omega\Delta-\delta-i\varepsilon}}
  \loga{\frac{\Delta+\delta-i\varepsilon}{\overline\mu}}
  +\loga{\frac{-(1-\Omega)\delta}{\Omega\Delta-\delta-i\varepsilon}}
  \loga{\frac{\Delta-\delta-i\varepsilon}{\overline\mu}}\right.\nonumber\\ 
& &   \left.
  -\Sp{\frac{\Delta+\delta-i\varepsilon}{\Delta-\delta/\Omega-i\varepsilon}}
  +\Sp{\frac{\Delta-\delta-i\varepsilon}{\Delta-\delta/\Omega-i\varepsilon}}
  + (\delta\rightarrow -\delta) \right\}~.  \label{m=0h}
\end{eqnarray}
Here, the terms within braces in the second and third line are to
be repeated with $\delta$ replaced by $-\delta$ as indicated.

When $\delta=0$, (\ref{m=0h}) reduces to (\ref{m=0=d}).  Thus, in the
case $m=0=\delta$ we have an analytic cross-check of our results.  In
the more general case $\delta\neq0$, we have numerically compared
general expression (\ref{finally}) for small values $m$, with
equations (\ref{m=0}) and (\ref{m=0h}).  The three expressions for
$\J{3}|_{\scsc m=0}$ were found to agree over a wide range of real
values for $\Delta_1$, $\Delta_2$, and $0<\Omega<1$, again providing
a cross-check for (\ref{finally}).

\section{The Scalar Two-Point Integral} 
\label{sec:2scalar}

The scalar two-point integral $\J{2}$ is given by (\ref{j2}) with
$n=0$.  $\J{2}$ is a function of $m$ and $\Delta=\delta M-2 v\cdot k$.
The starting point for the calculation of $\J{2}$ is (\ref{matchm3}).
The computations of both $\fJ{2}$ and $\J{2}|_{m=0}$ are standard.
Defining,
\begin{equation}
  \label{xpm}
  x_{\scsc\pm} = \frac{1}{2} \left(-\Delta\pm\sqrt{\Delta^2-4
  m^2+i\varepsilon}\right)~, 
\end{equation}
from (\ref{matchm3}) we obtain,
\begin{equation}
  \label{j2a}
  \J{2}  =  \frac{\Delta}{32\pi^2} \left(\frac{2}{\epsilon}+2\right) +
  \frac{1}{16\pi^2} \left(x_{\scsc +} \loga{-\frac{x_{\scsc +}}{\overline\mu}}
    +x_{\scsc -} \loga{-\frac{x_{\scsc -}}{\overline\mu}}\right)~.
\end{equation}
This result can also be obtained by using the HQET method for
combining denominators, which yields an equivalent expression in terms
of hypergeometric functions.

In order to compare our result (\ref{j2a}) for $\J{2}$ with those in
the previous literature, we rewrite it in terms of $\x=\Delta/(2m)$, 
\begin{equation}
  \label{j2b}
  \J{2}(\Delta,m)  =  \frac{\Delta}{32\pi^2} \left( \frac{2}{\epsilon} +
  \loga{\frac{\overline\mu^2}{m^2}} + 2 \right) + \frac{m}{16\pi^2}
  {\cal F}(\x) 
\end{equation}
with ${\cal F}(\x)  =  \sqrt{\x^2 - 1 + i\varepsilon} \left[
\loga{\x-\sqrt{\x^2 - 1 + i\varepsilon}} - \loga{\x+\sqrt{\x^2 - 1 +
i\varepsilon}}\right]$.  The coefficient of the dimensional
regularization pole vanishes when $\Delta=0$.  This is due to the
fact that the real part of the integrand in (\ref{j2}) is parity-odd
when $\Delta=0.$

Equation (\ref{j2b}) agrees with \cite{stew} for all real values of
$\Delta$, once their different normalization and conventions are taken
into account.  It agrees with the results from \cite{falk,boyd}
(see also the second of \cite{hhchpt}) only in the region $\x>0$, our
result being different from theirs over the entire negative semiaxis.

We consider now $\J{3}$ at zero recoil.  As shown in \cite{falk}, and
seen from its definition (\ref{j3}), $\J{3}$ at $\Omega=0$ can be
obtained from $\J{2}$ as,
\begin{equation}
  \label{j23}
 \left.\J{3}\rule{0ex}{2ex}\right|_{\scsc \Omega=0}(\Delta_1,
 \Delta_2) = \frac{1}{\Delta_1-\Delta_2} \left(\J{2}(\Delta_1,m) -
   \J{2}(\Delta_2,m)\right),
 \hspace{2ex}\mbox{and}\hspace{2ex}
\left.\J{3}\rule{0ex}{2ex}\right|_{\scsc \Omega=0}(\Delta,\Delta) =
\frac{\partial}{\partial\Delta} \J{2} (\Delta,m).
\end{equation}
Substituting the value of $\J{2}(\Delta,m)$ given by (\ref{j2b}) in
(\ref{j23}), we recover our previous results (\ref{j3zr}) and
(\ref{j3zrd0}), as can be easily checked. 

\section{Vector and Tensor Integrals}
\label{sec:tensor}

In this section we give general expressions for vector and second-rank
tensor integrals in terms of form factors.  We also compare our
results to those in the literature, when available.  The form factors
will be expressed in terms of scalar integrals.  Of those, $\J{2}$ and
$\J{3}$ have been given in previous sections.  We will also need,
\begin{equation}
  \label{j1}
  \J{1} = \frac{i\mu^{4-d}}{(2\pi)^d} 
  \int\! d^d\! q \frac{1}{\left(q^2-m^2+i\varepsilon\right)}
  = -\frac{m^2}{16\pi^2}
  \left(\frac{2}{\epsilon}+\loga{\frac{\overline\mu^2}{m^2}}+1\right)+
  {\cal O}(\epsilon)~,
\end{equation}
with $m>0$, $d=4-\epsilon$, $\loga{\overline\mu^2}=\loga{\mu^2
  4\pi}-\gamma_{\scsc\mathrm{E}}$.  Two other scalar integrals appear
in the evaluation of tensor ones,
\begin{equation}
  \hspace*{-1.5ex}
  \J{1}^\prime  =  \frac{i\mu^{4-d}}{(2\pi)^d} \int\! d^d\! q
  \frac{1}{\left(2v\!\cdot\! q-\Delta +i\varepsilon\right)}
  \hspace{1.5ex} \mbox{and} \hspace{1.5ex}
   \J{2}^\prime  =  \frac{i\mu^{4-d}}{(2\pi)^d}
  \int\! d^d\! q \frac{1}{
    \left(2v_1\!\cdot\! q-\Delta_1+i\varepsilon\right)
    \left(2v_2\!\cdot\! q-\Delta_2+i\varepsilon\right)} \label{j'12}
\end{equation}
Both $\J{1}^\prime$, and $\J{2}^\prime$ vanish, as we will now show.
The easiest way to see that $\J{1}^\prime=0$ is by applying the axioms
of dimensional regularization \cite[\S 4.1]{collins}.  We consider
$\J{1}^\prime (v^\mu,d)$ as a function of $d$ and $v^\mu$, momentarily
allowing $v^\mu v_\mu >0$, not necessarily equal to 1.  Then, we can
always shift the integration variable so that,
\[  \J{1}^\prime (v^\mu,d) = \frac{i\mu^{4-d}}{(2\pi^d)}\int\! d^d\! q
  \frac{1}{\left(2v\!\cdot\! q+i\varepsilon\right)}~. \]
Let $s>0$, and  consider $\J{1}^\prime (sv^\mu,d)$.  By factoring $s$
out of the integral we get, $\J{1}^\prime (sv^\mu,d)=1/s \J{1}^\prime
(v^\mu,d)$, whereas by rescaling the integration variable $q^\mu$ we
find, $\J{1}^\prime (sv^\mu,d)=1/s^d \J{1}^\prime (v^\mu,d)$.
Therefore, we must have $1/s \J{1}^\prime (v^\mu,d)=1/s^d \J{1}^\prime
(v^\mu,d)$ for all $s>0$ and all complex $d$, excluding positive
integer values.  Since $\J{1}^\prime$ must be an analytic function of
$d$, we conclude that it vanishes for $v^\mu v_\mu >0$.

We now turn to $\J{2}^\prime$.  Introducing a Feynman parameter, we
can write it as,
\[   \J{2}^\prime  =  \frac{i\mu^{4-d}}{(2\pi)^d} 
 \int_0^1\! dx 
 \int\! d^d\! q
  \frac{1}{2V(x)\!\cdot\! q-\Delta(x) +i\varepsilon}~,\]
with $V^\mu (x)=xv_1^\mu+(1-x)v_2^\mu$ and  $\Delta(x) = x\Delta_1 +
(1-x) \Delta_2$.  For $v_1^{\mu\,2}=1=v_2^{\mu\,2}$ and $0\leq x\leq
1$ we have $V^{\mu\,2}(x)>0$.  Thus, the inner integral is
 $\J{1}^\prime (V^\mu,d)$ and since $\J{1}^\prime = 0$,
 $\J{2}^\prime$ also vanishes. 

\subsection{Vector and tensor two-point integrals}
\label{sec:2vecten}

The vector two-point integral is given by,
\begin{equation}
  \label{2vec}
  \J{2}^{\mu\nu}(v^\alpha,\Delta,m) = \frac{i\mu^{4-d}}{(2\pi)^d} 
  \int\! d^d\! q \frac{q^\mu}{
    \left(2v\!\cdot\! q-\Delta+i\varepsilon\right)
    \left(q^2-m^2+i\varepsilon\right)}~.
\end{equation}
Lorentz invariance dictates that $\J{2}$ is given in terms of only one
form factor, which can be immediately evaluated by algebraic
reduction \cite{passar,denner},
\begin{equation}
  \label{2vec.a}
  \J{2}^\mu(v^\alpha,\Delta,m)  =  F(\Delta,m) v^\mu~,
  \hspace{2ex} \mbox{with} \hspace{2ex}  
  F(\Delta,m) =  v_\mu \J{2}^\mu(v^\alpha,\Delta,m) =
  \frac{1}{2} \J{1}(m) + \frac{\Delta}{2}\J{2}(\Delta,m)~,
\end{equation}
where the scalar integrals $\J{1}$ and $\J{2}$ have been given in
(\ref{j1}) and in section \ref{sec:2scalar}, respectively.

The tensor two-point integral is defined as,
\begin{equation}
  \label{2ten}
  \J{2}^{\mu\nu}(v^\alpha,\Delta,m) = \frac{i\mu^{4-d}}{(2\pi)^d}
  \int\! d^d\! q \frac{q^\mu q^\nu}{
    \left(2v\!\cdot\! q-\Delta+i\varepsilon\right)
    \left(q^2-m^2+i\varepsilon\right)}.
\end{equation}
We will introduce two sets of form factors.  First, we define,
\begin{equation}
  \label{2ten.a}
  \J{2}^{\mu\nu}(v^\alpha,\Delta,m) = I_0(\Delta,m) g^{\mu\nu} +
  I_1(\Delta,m) v^\mu v^\nu.
\end{equation}
Second, we introduce $F$ form factors which can be easily 
computed in terms of scalar integrals,
\begin{eqnarray}
  \label{2ten.b}
  F_0(\Delta,m) & \equiv & g_{\mu\nu} \J{2}^{\mu\nu} = d I_0 + I_1 
   =  m^2 \J{2}(\Delta,m)\\
  F_1(\Delta,m) &\equiv & v_\mu v_\nu \J{2}^{\mu\nu} = I_0+I_1
   =  \frac{\Delta}{4} \left(\J{1}(m)+\Delta
    \J{2}(\Delta,m)\rule{0ex}{2ex}\right)~.
\end{eqnarray}
In fact, $F_0(\Delta,m)=\J{1}^\prime+m^2 \J{2}(\Delta,m)$, so
here we have used $\J{1}^\prime=0$.  Inverting the relation among
$F$'s and $I$'s we obtain, to lowest order in $\epsilon=4-d$,
\begin{eqnarray}
  \label{2ten.c}
  I_0(\Delta,m) & = & -\frac{1}{3} \left(1+\frac{\epsilon}{3}\right)
  \left[ \frac{\Delta}{4}\J{1}(m)+\left(\frac{\Delta^2}{4}-m^2\right)
    \J{2}(\Delta,m)\right]\\
  I_1(\Delta,m) & = & \frac{\Delta}{3} \left(1 + \frac{\epsilon}{12}
  \right) \left[\J{1}(m)+\Delta \J{2}(\Delta,m)\rule{0ex}{2ex}\right]
  - \frac{m^2}{3} \left(1+\frac{\epsilon}{3}\right) \J{2}(\Delta,m)~.
\end{eqnarray}
Finally, we substitute the known values of $\J{1}(m)$ and
$\J{2}(\Delta,m)$.  Using the same notation as in (\ref{j2b}),
\begin{eqnarray}
  \label{2ten.d}
  I_0(\Delta,m) & = & -\frac{m^3}{3\cdot 16\pi^2}
  \left\{\left(\frac{2}{\epsilon} + \loga{\frac{\overline\mu^2}{m^2}}
  + \frac{8}{3}\right) \x \left(\x^2-\frac{3}{2}\right) + \frac{\x}{2}+
  (\x^2-1) {\cal F}(\x)\right\} \\
  \label{2ten.e}
  I_1(\Delta,m) & = & \frac{m^3}{3\cdot 16\pi^2}
  \left\{\left(\frac{2}{\epsilon} + \loga{\frac{\overline\mu^2}{m^2}}
  + \frac{13}{6}\right) \x (4 \x^2-3) + \frac{3}{2} \x + (4\x^2 - 1)
  {\cal F}(\x) \right\}~.
\end{eqnarray}
This is the general form for $\J{2}$.  We do not find agreement
with \cite{boyd}.  

There are two particular cases of interest, in which $\J{2}^{\mu\nu}$
can be easily computed directly by using the HQET method for combining
denominators, thus providing cross-checks for our results.  In the
first place,  we consider the case $m=0$, $\Delta>0$ (the case
$\Delta<0$ is analogous).  A straightforward computation using the
HQET method yields,
\begin{equation}
  \label{2ten.m0}
  \left. I_0\rule{0ex}{2ex}\right|_{\stackrel{\scsc m=0}{\scsc
  \Delta>0}}  =   - \frac{\Delta^3}{3\cdot 128\pi^2}
  \left(\frac{2}{\epsilon} + \loga{\frac{\overline\mu^2}{\Delta^2}} +
  \frac{8}{3}\right)~;
  \hspace{3ex}
  \left. I_1\rule{0ex}{2ex}\right|_{\stackrel{\scsc m=0}{\scsc
  \Delta>0}}  =   \frac{\Delta^3}{3\cdot 32\pi^2}
  \left(\frac{2}{\epsilon} + \loga{\frac{\overline\mu^2}{\Delta^2}} +
  \frac{13}{6}\right)
\end{equation}
which agree with (\ref{2ten.d}) and (\ref{2ten.e}) evaluated at $m=0$.

Second, in the case $\Delta=0$ we find,
\begin{equation}
  \label{2ten.d0}
  \J{2}^{\mu\nu}(v^\alpha,\Delta=0,m) = \frac{m^3}{3\cdot 16\pi}
  (g^{\mu\nu} - v^\mu v^\nu)~,
\end{equation}
again in agreement with the corresponding limit of (\ref{2ten.d}) and
(\ref{2ten.e}).  As remarked above, in connection with the scalar
integral, there is no dimensional regularization pole in this case.

Furthermore, if we assume $0<\x<1$ ($m > \Delta/2 > 0$) and expand in
powers of $\x$, we recover the result given in \cite{cho}.

\subsection{Vector three-point integral}
\label{sec:3vec}

We now turn to the tensor three-point integrals, starting with the
vector one,
\begin{equation}
  \label{3vec}
  \J{3}^\mu (v_1^\alpha, v_2^\beta, \Delta_1, \Delta_2, m) =
  \frac{i\mu^{4-d}}{(2\pi)^d}   \int\! d^d\! q \frac{q^\mu}{
    \left(2v_1\!\cdot\! q-\Delta_1+i\varepsilon\right)
    \left(2v_2\!\cdot\! q-\Delta_2+i\varepsilon\right)    
    \left(q^2-m^2+i\varepsilon\right)}.
\end{equation}
On the left-hand side we omitted $\mu$ and $d$ from the argument for
brevity.  We define, as before, two sets of form factors,
\begin{equation}
  \label{3vec.a}
  \J{3}^\mu = I_1 v_1^\mu + I_2 v_2^\mu
  \hspace{3ex}\mbox{and}\hspace{3ex}
  F_{1,2} =  v_{1,2}\cdot\J{3}~,
\end{equation}
with $I_j (\Omega,\Delta_1,\Delta_2,m)$ and $F_j
(\Omega,\Delta_1,\Delta_2,m)$ related by,
\begin{equation}
  \label{3vec.b}
  I_1  =  \frac{1-\Omega^2}{4\Omega^2} \left[-(1-\Omega^2) F_1 +
    (1+\Omega^2) F_2 \right]~,
  \hspace{3ex}
  I_2  =  \frac{1-\Omega^2}{4\Omega^2} \left[(1+\Omega^2) F_1 -
  (1-\Omega^2) F_2 \right] ~.
\end{equation}
The form factors $F_j$ can be expressed in terms of scalar integrals
as,
\begin{equation}
  \label{3vec.c}
  F_{1,2} = \frac{1}{2} \J{2}(\Delta_{2,1},m) + \frac{\Delta_{1,2}}{2}
  \J{3}(\Omega,\Delta_1,\Delta_2,m)~. 
\end{equation}
These equations give an explicit expression for $\J{3}^\mu$.

At zero recoil we have $v_1^\mu = v_2^\mu = v^\mu$ and,
\begin{equation}
  \label{3vec.zr}
  \left.\J{3}^\mu\rule{0ex}{2ex}\right|_{\Omega=0}  =  v_\nu\cdot
  \left.\J{3}^\nu\rule{0ex}{2ex}\right|_{\Omega=0} v^\mu~,
  \hspace{1ex}\mbox{with}\hspace{2ex}
  v_\nu\cdot \left.\J{3}^\nu\rule{0ex}{2ex}\right|_{\Omega=0}  = 
  \frac{1}{2}\J{2}(\Delta_2,m) + \frac{\Delta_1}{2}
  \J{3}(\Omega=0,\Delta_1,\Delta_2,m)~.
\end{equation}
Using (\ref{j23}) we obtain,
\[v_\nu\cdot \left.\J{3}^\nu\rule{0ex}{2ex}\right|_{\Omega=0} = 
\frac{1}{2} \frac{1}{\Delta_1-\Delta_2} \left[ \Delta_1
  \J{2}(\Delta_1,m) - \Delta_2 \J{2}(\Delta_2,m)\right]\]
if $\Delta_1 \neq \Delta_2$ and
\[v_\nu\cdot \left.\J{3}^\nu\rule{0ex}{2ex}\right|_{\Omega=0} = 
\frac{1}{2} \J{2}(\Delta,m) + \frac{\Delta}{2}
\frac{\partial}{\partial\Delta} \J{2}(\Delta,m)\]
if $\Delta_1 = \Delta_2 = \Delta$.  This completes our treatment of
the vector integral.

\subsection{The tensor three-point integral}
\label{sec:3ten}

We consider, finally, the tensor integral,
\begin{equation}
  \label{3ten}
  \J{3}^{\mu\nu} (v_1^\alpha, v_2^\beta, \Delta_1, \Delta_2, m) =
  \frac{i\mu^{4-d}}{(2\pi)^d}   \int\! d^d\! q \frac{q^\mu q^\nu}{
    \left(2v_1\!\cdot\! q-\Delta_1+i\varepsilon\right)
    \left(2v_2\!\cdot\! q-\Delta_2+i\varepsilon\right)    
    \left(q^2-m^2+i\varepsilon\right)}.
\end{equation}
In this case we have two sets of four form-factors each,
\begin{equation}
  \label{3ten.a}
  \J{3}^{\mu\nu} = I_{11} v_1^\mu v_1^\nu + I_{22} v_2^\mu v_2^\nu +
  I_{12} v_1^{\{\mu} v_2^{\nu\}} + I_{0} g^{\mu\nu}~,
\end{equation}
with $v_1^{\{\mu} v_2^{\nu\}} = v_1^\mu v_2^\nu + v_2^\mu v_1^\nu$,
and, 
\begin{equation}
  \label{3ten.b}
  F_{11} = v_1^\mu v_1^\nu \J{3}_{\mu\nu}~;\hspace{2ex}
  F_{22} = v_2^\mu v_2^\nu \J{3}_{\mu\nu}~;\hspace{2ex}
  F_{12} = v_1^{\{\mu} v_2^{\nu\}} \J{3}_{\mu\nu}~;\hspace{2ex}
  F_{0} = g^{\mu\nu} \J{3}_{\mu\nu}~.
\end{equation}
The $F$'s can be given in terms of the $I$'s using (\ref{3ten.a}).
Inverting those relations we obtain,
\begin{eqnarray}
  I_{11} & = & \frac{d-1}{d-2} \frac{(1-\Omega^2)^2}{16\Omega^4}
  \left\{(1-\Omega^2)^2 F_{11}+\left(\Omega^4 + 2 \frac{d-3}{d-1}
  \Omega^2 + 1\right) F_{22} - (1-\Omega^4) F_{12} \right\}\nonumber\\
   &  & + \frac{1}{d-2}  \frac{(1-\Omega^2)^2}{4 \Omega^2} F_0\nonumber\\
  I_{22} & = & \frac{d-1}{d-2} \frac{(1-\Omega^2)^2}{16\Omega^4}
  \left\{\left(\Omega^4+2 \frac{d-3}{d-1} \Omega^2+1\right) F_{11}+
    (1-\Omega^2)^2 F_{22}-(1-\Omega^4) F_{12} \right\}\nonumber\\
  &  & + \frac{1}{d-2}  \frac{(1-\Omega^2)^2}{4 \Omega^2} F_0  \label{3ten.c}\\
  I_{12} & = & -\frac{d-1}{d-2} \frac{(1-\Omega^2)^2}{16\Omega^4}
  \left\{(1-\Omega^4) (F_{11}+F_{22}) - \left(\Omega^4 +
      \frac{2}{d-1}\Omega^2 + 1\right) F_{12} \right\}\nonumber\\
  & &  - \frac{1}{d-2} \frac{1-\Omega^4}{4\Omega^2} F_0\nonumber\\
  I_0 & = & \frac{1}{d-2} \frac{(1-\Omega^2)^2}{4\Omega^2}
  (F_{11}+F_{22}) + \frac{1}{d-2} F_0 -\frac{1}{d-2}
  \frac{1-\Omega^4}{4\Omega^2} F_{12}~.\nonumber
\end{eqnarray}
Using the results from sections \ref{sec:2vecten} and \ref{sec:3vec}
we can express the $F$'s in terms of scalar integrals as,
\begin{eqnarray}
  F_{11} & = & \frac{\omega}{4} \J{1}(m) +
  \frac{\Delta_1+\omega\Delta_2}{4} \J{2}(\Delta_2,m) +
  \frac{\Delta_1^2}{4} \J{3}(\Omega,\Delta_1,\Delta_2,m) \nonumber\\
  F_{22} & = & \frac{\omega}{4} \J{1}(m) +
  \frac{\Delta_2+\omega\Delta_1}{4} \J{2}(\Delta_1,m) +
  \frac{\Delta_2^2}{4} \J{3}(\Omega,\Delta_1,\Delta_2,m)   \label{3ten.d}\\
  F_{12} & = & \frac{1}{2} \J{1}(m) + \frac{1}{2} \left(\Delta_1
  \J{2}(\Delta_1,m) + \Delta_2 \J{2}(\Delta_2,m)\rule{0ex}{2ex}\right) +
  \frac{\Delta_1\Delta_2}{2} \J{3}(\Omega,\Delta_1,\Delta_2,m) \nonumber\\
  F_0 & = & m^2 \J{3}(\Omega,\Delta_1,\Delta_2,m)~.\nonumber
\end{eqnarray}
Here we are using a mixed notation, in terms of both $\omega$ and
$\Omega$ (see (\ref{omega})), for brevity.  Equations (\ref{3ten.a}),
(\ref{3ten.c}) and (\ref{3ten.d}) give an explicit analytic expression
for $\J{3}^{\mu\nu}$.  Notice also the symmetry of (\ref{3ten.d})
under exchange of $\Delta_1$ and $\Delta_2$.  The same as with the
scalar integral, there are a number of particular cases of interest
which we briefly comment upon in the remainder of this section.

\subsubsection{The zero recoil case}

In order to study the zero recoil case, it is convenient to write
$\J{3}^{\mu\nu}$ in terms of vectors $v_\pm^\mu = 1/2 (v_1^\mu \pm
v_2^\mu)$.  Instead of (\ref{3ten.a}) we then have,
\begin{equation}
  \label{3ten.e}
    \J{3}^{\mu\nu} = I_{++} v_+^\mu v_+^\nu + I_{--} v_-^\mu v_-^\nu +
  I_{+-} v_+^{\{\mu} v_-^{\nu\}} + I_{0} g^{\mu\nu}~.
\end{equation}
In the zero recoil limit, $v_1^\mu = v_2^\mu = v_+^\mu \equiv v^\mu$.
Using the results of sections \ref{sec:2vecten} and \ref{sec:3vec} it
is not difficult to show that when $\Omega=0$ we have $v_-^\mu
\J{3}_{\mu\nu}=0$. Therefore, we can write,
\begin{equation}
  \label{3ten.f}
  \left.\J{3}^{\mu\nu}\rule{0ex}{2ex}\right|_{\Omega=0} =
  \left.I_{++}\rule{0ex}{2ex}\right|_{\Omega=0} v^\mu v^\nu +
  \left.I_{0}\rule{0ex}{2ex}\right|_{\Omega=0} g^{\mu\nu}~.
\end{equation}
These form factors can be computed as before, resulting in,
\begin{eqnarray}
  \label{3ten.g}
  \left.I_{++}\rule{0ex}{2ex}\right|_{\Omega=0} & = & 
  \frac{1}{3}\left(1+\frac{\epsilon}{12}\right) \left\{ \J{1}(m) +
  \frac{\Delta_1^2 - m^2}{\Delta_1 - \Delta_2}  \J{2}(\Delta_1,m) - 
  \frac{\Delta_2^2 - m^2}{\Delta_1 - \Delta_2}  \J{2}(\Delta_2,m)
  )\right\} \\ 
  \left.I_{0}\rule{0ex}{2ex}\right|_{\Omega=0} & = & 
  -\frac{1}{12}\left(1+\frac{\epsilon}{3}\right) \left\{ \J{1}(m) +
  \frac{\Delta_1^2 - 4 m^2}{\Delta_1 - \Delta_2}  \J{2}(\Delta_1,m) - 
  \frac{\Delta_2^2 - 4 m^2}{\Delta_1 - \Delta_2}  \J{2}(\Delta_2,m)
  )\right\}~.
\end{eqnarray}
We notice that we could have arrived at these equations by using a
relation analogous to (\ref{j23}), namely, 
\begin{equation}
  \label{3ten.h}
\left.\J{3}^{\mu\nu}\rule{0ex}{2ex}\right|_{\scsc \Omega=0}
= \frac{1}{\Delta_1-\Delta_2}
\left(\J{2}^{\mu\nu}(v^\alpha,\Delta_1,m) -
\J{2}^{\mu\nu}(v^\alpha,\Delta_2,m)\right)~,  
\end{equation}
showing the consistency of our result.  These expressions acquire a
particularly simple form for some special values of the parameters.
For instance, setting $m=0$, $\Delta_j >0$, $j=1,2$ and using either
(\ref{3ten.h}) and  (\ref{2ten.m0}),  or (\ref{3ten.g}), we get,
\begin{eqnarray}
  \label{3ten.zr.m0.a}
  \left. I_0\rule{0ex}{2ex}\right|_{\stackrel{\scsc \Omega=0}{\scsc
  m=0}}& = & \frac{-1}{3\cdot 128\pi^2} \frac{1}{\Delta_1-\Delta_2}
   \left\{\Delta_1^3 \left(\frac{2}{\epsilon} +
   \loga{\frac{\overline\mu^2}{\Delta_1^2}} + \frac{8}{3}\right) -
   \Delta_2^3 \left(\frac{2}{\epsilon} +
   \loga{\frac{\overline\mu^2}{\Delta_2^2}} + \frac{8}{3}\right)
   \right\} \\
    \left. I_{++}\rule{0ex}{2ex}\right|_{\stackrel{\scsc \Omega=0}{\scsc
  m=0}}& = & \frac{1}{3\cdot 32\pi^2} \frac{1}{\Delta_1-\Delta_2}
   \left\{\Delta_1^3 \left(\frac{2}{\epsilon} +
   \loga{\frac{\overline\mu^2}{\Delta_1^2}} + \frac{13}{6}\right) -
   \Delta_2^3 \left(\frac{2}{\epsilon} +
   \loga{\frac{\overline\mu^2}{\Delta_2^2}} + \frac{13}{6}\right)
   \right\},
\end{eqnarray}
which provides another cross-check of our previous equations.

\subsubsection{The case $\Delta_1 = 0 = \Delta_2$}

Another case where the form factors for $\J{3}^{\mu\nu}$ take a very
simple form is when $\Delta_1 = 0 = \Delta_2$.  In this case equations
(\ref{3ten.a}), (\ref{3ten.c}) and (\ref{3ten.d}), together with our
previous results for the scalar integrals, give,
\begin{eqnarray}
  \left. I_0\rule{0ex}{2ex}\right|_{\Delta_j=0} & = &
  \frac{m^2}{128\pi^2} \frac{1-\Omega^2}{\Omega} \loga{\alp}
  \left(\frac{2}{\epsilon} + \loga{\frac{\overline\mu^2}{m^2}} +
  1\right)\nonumber\\
  \left. I_{11}\rule{0ex}{2ex}\right|_{\Delta_j=0} & = &
  \frac{m^2}{128\pi^2} \left[-\frac{1-\Omega^4}{2\Omega^2} +
    \frac{(1-\Omega^2)^3}{4 \Omega^3} \loga{\alp}\right]
  \left(\frac{2}{\epsilon} + \loga{\frac{\overline\mu^2}{m^2}} +
    1\right)\nonumber\\
  \left. I_{22}\rule{0ex}{2ex}\right|_{\Delta_j=0} & = &
  \left. I_{11}\rule{0ex}{2ex}\right|_{\Delta_j=0}  \label{3ten.dd0}\\
  \left. I_{12}\rule{0ex}{2ex}\right|_{\Delta_j=0} & = &
  \frac{m^2}{128\pi^2} \left[\frac{(1-\Omega^2)^2}{2\Omega^2} -
    \frac{(1-\Omega^2)(1-\Omega^4)}{4 \Omega^3} \loga{\alp}\right]
  \left(\frac{2}{\epsilon} + \loga{\frac{\overline\mu^2}{m^2}} +
    1\right)~.\nonumber
\end{eqnarray}
We have also computed $\J{3}^{\mu\nu}$ for $\Delta_j=0$ directly from
its definition (\ref{3ten}) by using the HQET method for combining
denominators.  Full agreement with (\ref{3ten.dd0}) was found.

\subsubsection{The case $m=0$}

A direct calculation of $\J{3}^{\mu\nu}$ at $m=0$ with the HQET method
is considerably more involved than in the previous case.  The results
are also much less compact.  As an illustration, we will quote the
result for the form factor $I_0$ when $m=0$ and $\Delta_1 = \Delta_2
\equiv \Delta >0$, 
\begin{eqnarray} 
  \label{3ten.m0}
  I_0  & = & -\frac{\Delta^2}{256\pi^2} (1-\Omega^2) \left\{ \left(1 +
  \frac{1-\Omega^2}{2\Omega} \loga{\alp} \right) \left(
  \frac{2}{\epsilon} + \loga{\frac{\overline\mu^2}{\Delta^2}}\right)
  + 3 + \frac{(1-\Omega^2)}{2\Omega} \loga{\alp} \right.\nonumber\\
  &  & \left.
   - \frac{(1-\Omega^2)}{4\Omega} \loga{\frac{1-\Omega^2}{4}} \loga{\alp}
  + \frac{(1-\Omega^2)}{2\Omega}
  \left(\Sp{\frac{1-\Omega}{2}} - \Sp{\frac{1+\Omega}{2}}\right)\right\}~.
\end{eqnarray}
This equation agrees with the general result given by (\ref{3ten.c})
and (\ref{3ten.d}), evaluated at $m=0$,  $\Delta_1 = \Delta_2$, as it
should. 

\subsubsection{The chiral logs}

In this section we focus on the case $\Delta_1 = \Delta_2 \equiv \delta m$,
with $0 < \delta m/2 < m$.  We expand in powers of $\delta m/m$,
retaining only those terms proportional to $\loga{m}$, with coefficients
computed to lowest order in $\delta m/m$.  In this way, we obtain the
chiral logs in $\J{3}^{\mu\nu}$.

From equations (\ref{j2b}) and (\ref{j1}), we get,
\begin{equation}
  \label{log.a}
  \J{1}(m)  =  -\frac{m^2}{16\pi^2} \loga{\frac{\overline\mu^2}{m^2}} +
  \cdots~;
  \hspace{3ex}
  \J{2}(\delta m,m)  =  \frac{\delta m}{32\pi^2}
  \loga{\frac{\overline\mu^2}{m^2}} + \cdots~,
\end{equation}
where the ellipsis denotes terms not containing $\loga{m}$, or
containing higher powers of $\delta m$.  On the other hand, $\J{3}$ is
needed only to zeroth order in $\delta m$ because it enters the form
factors with $\delta m^2$ as a coefficient.  From (\ref{innermost}),
\begin{equation}
  \label{log.b}
  \J{3}(\Omega,0,0,m) = \frac{1}{64\pi^2} \frac{1-\Omega^2}{\Omega}
  \loga{\alp} \loga{\frac{\overline\mu^2}{m^2}} + \cdots~.
\end{equation}
With these approximations, we obtain the form factors as, 
\begin{eqnarray}
  F_{11} = F_{22} & = & \frac{1}{64\pi^2} \loga{\frac{\overline\mu^2}{m^2}}
  \left\{-\frac{1+\Omega^2}{1-\Omega^2} m^2 + \delta m^2 \left(
    \frac{1}{1-\Omega^2} + \frac{1-\Omega^2}{4\Omega}
  \loga{\alp}\right) \right\} + \cdots \\
  F_{12} & = & \frac{1}{64\pi^2}\loga{\frac{\overline\mu^2}{m^2}}
  \left\{-2 m^2 + \delta m^2 \left(2 + \frac{1-\Omega^2}{2\Omega}
      \loga{\alp} \right)\right\} + \cdots \\
  F_{0} & = & \frac{1}{64\pi^2}\loga{\frac{\overline\mu^2}{m^2}}
  m^2 \frac{1-\Omega^2}{\Omega} \loga{\alp} + \cdots~.
\end{eqnarray}
These results agree exactly with those of \cite{cho}, once we take into
account the differences in normalization and  conventions.

\section{Final Remarks}
\label{sec:final}

In phenomenological applications, the exact functional dependence of
Feynman integrals on masses and residual momenta is usually not
needed.  Often, the first few terms in a series expansion in some of
the parameters provides the required accuracy.  We believe, however,
that the exact analytic computation presented here does not require
more calculational effort than approximate schemes.  It has the added
advantage of being valid over the entire physical region for internal
and external masses.

Our result (\ref{finally}) for the scalar three-point integral
involves four dilogarithms.  This is to be compared with the analogous
vertex integrals in renormalizable theories, which are generally
expressed in terms of twelve dilogarithms and a collection of
logarithms \cite{thooft,denner}.  This simplification is afforded, of
course, by the effective theory formalism, which focuses only on the
relevant degrees of freedom.  Equation (\ref{finally}) is quite
compact.  Once the values for internal and external masses are given,
so that the appropriate branches of square roots, logs and dilogs are
determined with the aid of the ``$i\varepsilon$'' prescription, the
expression for $\J{3}$ given by (\ref{finally}) is easily translated
into computer code.

Another possible approach to the computation of three-point integrals
is to consider them strictly within the context of the effective
theory, without introducing auxiliary integrals such as $\fJ{3}$, eq.\ 
(\ref{fj3}).  In that case, one can parametrize the integrand with the
HQET method.  The resulting expressions are, however, difficult to
handle and, in general, they seem to lead to hypergeometric functions
of two variables or, more likely, to series of hypergeometric
functions.  The procedure adopted in this paper avoids those
difficulties.

\section*{Acknowledgements}

I would like to thank my colleagues at CINVESTAV-M\'erida for their
help, and especially Prof.\ V.Gupta for discussions.

\noindent This work was partially supported by Conacyt and SNI.

\end{document}